\documentclass[pre,aps,floats,superscriptaddress,floatfix,twocolumn]{revtex4}
\usepackage{amssymb,amsmath}
\usepackage{amsmath,amssymb}
\usepackage{graphicx}
\usepackage{psfrag}
\usepackage{color}

\def\beq{\begin{equation}}
\def\eeq{\end{equation}}
\def\bea{\begin{eqnarray}}
\def\eea{\end{eqnarray}}
\begin{document}
 %\title{  Order and synchronization of two-dimensional chemical 
%oscillators by 
%active catalysis}
%\title{Phase 
%fluctuations in a collection of active oscillators}
%\title{Synchronization and ordering of oscillators in an active medium}
\title{Active hydrodynamics of synchronization and ordering in moving 
oscillators}
\author{Tirthankar Banerjee}\email{tirthankar.banerjee@saha.ac.in}
\author{Abhik Basu}\email{abhik.basu@saha.ac.in,abhik.123@gmail.com}
\affiliation{Condensed Matter Physics Division, Saha Institute of
Nuclear Physics, Calcutta 700064, India}

\date{\today}
\begin{abstract}

%What can control synchronization of coupled oscillators? 
 The nature of emergent collective behaviors of moving physical 
agents interacting with their neighborhood is a
long-standing 
open issue in
physical and biological systems alike. This calls for studies on the
control of synchronization and the degree of order in a collection
of diffusively moving noisy oscillators.  We address this by 
constructing 
 a generic hydrodynamic theory for {\em active} phase fluctuations in 
 a collection of large number of nearly phase-coherent moving oscillators in
 two dimensions. Our theory { describes the general situation} where 
phase 
fluctuations and oscillator
mobility mutually affect each other.
  We show that the interplay between the active effects and the
mobility of the oscillators
leads to a variety of phenomena, ranging from synchronization with long range, 
nearly long range and quasi long 
range orders to instabilities and 
desynchronization with short range order of the oscillator phases. 
We highlight the complex dependences of synchronization on 
the active effects. These should be testable in wide ranging systems, 
e.g., oscillating
chemical reactions in the presence of different reaction 
inhibitors/facilitators,  live oriented cytoskeletal extracts, or vertebrate 
segmentation clocks. 
\end{abstract}

\maketitle
%{\em Introduction:-}
\section{Introduction}

 The phenomenon of synchronization in which a large number of 
microscopic units spontaneously organize themselves into displaying cooperative 
behavior plays an important role in a wide class of systems, ranging from physics 
and biology to ecology, social dynamics and 
neurosciences~\cite{pikovsky-book,strogatz-book}. Cooperative 
behavior in many
living systems made of large number of living beings can be observed across a 
range of 
biological systems, e.g, a suspension of cells synchronizing their genetic 
clocks~\cite{cells1,cells2,cells3,cells4,cells5} and flashing of fire 
flies~\cite{fire-fly}, and their artificial imitations~\cite{arti-imi}. For 
instance, when genetic oscillators that control the expression of a  
fluorescent 
protein is inserted in {\em E. Coli} bacteria, they can flash at a regular 
rate~\cite{cells1,ecoli}; when coupled, a large population of bacteria can 
flash 
rhythmically in a synchronized manner~\cite{cells2,cells4}. Two other notable 
relevant examples are vertebrate segmentation 
clock~\cite{vertebrate} and oscillating chemical reactions~\cite{stirred-BZ}.

The collective excitations in a large number of these 
nonequilibrium 
physical, chemical,
and biological systems are in the form of cooperative oscillations of active 
interacting
elements~\cite{strogatz}, e.g., 
chemical oscillators~\cite{miyakawa-pre2002,toiya1,toiya2, 
fukuda-jphys, 
kiss-prl,migliorini, strogatz-physicad,BZ-vortical} and synthetic genetic 
oscillators~\cite{gen-osc}, and biologically relevant 
systems~\cite{frank22,frank-chiral,sebastian,liverpool,
frank-exp,somitogenesis,deve,quorum,Laskar}.  Spontaneous locking of 
interacting oscillators to a common phase~\cite{dofler} lead to
{\em synchronization} of the oscillators, a ubiquitous collective 
behavior~\cite{pikovsky-book,strogatz-book}, {observed, e.g.,
in complex networks of phase oscillators~\cite{dofler,net1}}. These call for 
studies on 
synchronization of a collection of {\em locally interacting
mobile} 
 oscillators.  These have 
been studied extensively, in particular, in agent-based discrete 
systems~\cite{frasca,frank2,lauter,pagona}, where the network usually consists 
of 
a group of interacting moving oscillators. Equivalently, these 
model studies may also be viewed as
examples of synchronization in dynamical networks, where the connectivity 
between any two oscillators evolves in time~\cite{pagona,chaos,net1}.
 Recent studies  on agent-based models with a large number of interacting 
mobile 
oscillators
 in one (1d) and two (2d) dimensions indicate that increasing 
the mobility of the oscillators significantly affect the steady states and may 
even lead to global 
synchronization~\cite{frasca,peruani}. In addition, recent works provide 
evidence in favor of cell movement promoting 
synchronization of coupled genetic oscillators~\cite{uriu1,uriu2}. Our 
work complement these existing studies. General 
understanding of how 
mobility of the oscillators affects synchronization (or, 
lack thereof) in a collection of 
mobile oscillators form the principal motivation of this work.

%The 
%associated theoretical descriptions at the simplest level are often made
%in terms of
%interacting oscillators~\cite{sebastian,frank-risler}.
%Nonequilibrium analogs of broken symmetry hydrodynamics for 
%equilibrium ordered phases~\cite{martin} have
%recently been successfully 
%applied to spatially extended ordered {\em active} systems with 
%continuous symmetries~\cite{sriram-rmp,active-smectics}. 

 In this article, we focus on the 
the generic long wavelength
properties of small fluctuations in a collection of 
 diffusively mobile, 
nearly phase-coherent, noisy out-of-equilibrium oscillators in 2d. We
introduce a generic active hydrodynamic theory  for such systems.  
 We analyze the broken 
symmetry  phase 
fluctuations of these oscillators in their nearly phase coherent states and 
examine the general conditions for synchronization.  Hydrodynamic 
approaches are distinguished by their generality of predictions and have been 
successfully applied to the ordered broken symmetry phases of many equilibrium 
and 
nonequilibrium systems~\cite{martin-parodi-pershan,active-fluid}. Hydrodynamic 
theories are particularly suitable to extract statistical properties in the 
limit of large distance and long time 
scales~~\cite{martin-parodi-pershan,active-fluid}. In order to generalize the 
scope of our study, we study a collection of large number of 
diffusively mobile oscillators in the continuum limit, where the local phase 
and number density fluctuations can mutually affect each other. 
In other words, in a discrete, agent-based description, the agents undergo 
persistent random walk that depends on the local phase 
fluctuations~\cite{affect}.
The active 
interplay between the oscillator phases and the oscillator mobility is 
shown to 
control the 
degree of phase coherence. This forms the principal result of this
work.
Our model provides a generic long wavelength 
description for {\em active mobility-induced 
synchronization}~\cite{frasca,peruani,markus,mobility} in 2d  with 
the additional feature that the mobility is affected by local phase 
fluctuations. We expect it 
to be relevant in experiments pertaining to a wide class of systems, ranging 
from oscillating chemical reactions~\cite{b-z,b-r,b-l,i-c} in the presence of 
catalysts to vertebrate segmentation clocks~\cite{vertebrate,deve,frank2}, 
oriented live
cytoskeletal extracts~\cite{sebastian} and clock 
synchronization in mobile robots~\cite{chaos} as well as help in designing new
artificial imitations~\cite{arti-imi}.  Our work 
generalizes studies on synchronization in complex dynamical 
network~\cite{frasca,pagona,peruani,net1,affect}, where the 
time-evolution of the network is affected by the oscillator phases.
In order to concentrate on the essential physics of  phase fluctuations and 
ordering~\cite{lauter}, we study the active stochastic dynamics of a collection 
of diffusive particles of concentration $c({\bf x},t)$
in 2d on a rigid substrate, each carrying
 oscillators  in their nearly 
phase-coherent state. The oscillators are 
represented 
by a complex field 
$Z=\exp[i\phi({\bf x},t)]$ with unit amplitude and phase $\phi$ at point $\bf 
x$ and time $t$, and have the same internal symmetry as the XY 
model~\cite{classicalxy}.    In 
stark contrast to related 2d equilibrium systems with XY 
symmetry,
we show that this model 
displays a wide class of 
behaviors, ranging from synchronization with
long range order (LRO), quasi-long range order (QLRO) and nearly long range 
order (NLO) to linear instability with desynchorinization and short range order 
(SRO),
and nonlinear 
stabilization 
of linear instability.   These are controlled by the interplay between 
active effects and particle diffusivity. In analogy with thermally 
excited systems, these regimes with 
different natures of order are characterized by the analog of the {\em 
Debye-Waller factor} $\sim \exp(-\Delta)$, where  
$\Delta=\langle \phi({\bf x},t)^2\rangle$ ($\langle...\rangle$ 
implies averages over the noises)~\cite{classicalxy,o_n,comment-debye}.  Here, 
$\Delta$  provides a measure of 
order or degree 
of phase synchronization: for a given $L$, where $L$ is the system size,
the smaller $\Delta$ is, the higher is the order or the degree of synchronization.
In particular,  in the linearly 
stable regime and in the 
limit  of fast concentration relaxation (also called {\em fast switching 
regime}, see below), $\Delta$
 can be reduced by enhancing 
(positive) $\gamma$, the 
active damping of phase fluctuations, 
with either LRO (finite $\Delta$) or NLO 
($\Delta$ varying as $\ln\ln L$) in the system. 
%becoming finite (hence LRO) {\bf in TL, i.e., linear system size 
%$L\rightarrow \infty$, 
%For intermediate $D_c$, 
%depending upon other details, $\Delta$ may scale as $\ln L$ or $\ln\ln L$. 
For 
negative $\gamma$, linear instability ensues, implying  SRO or
desynchronization.  Additionally, in some cases formation of patterns are 
predicted. Our work reveals complex dependences of 
phase fluctuations of a collection of moving oscillators on active effects.
 The rest of the article is organized as follows: In Sec.~\ref{model} we
construct our model and describe the origin of the active terms. Then in Sec.~\ref{result}
we discuss our results for both the linear and nonlinear theories. Finally, in Sec.~\ref{summ} we
summarize and conclude. An Appendix containing some calculational details has 
been added at the end for helping the 
readers.

\section{Construction of the model equations}\label{model}

{ We consider small fluctuations about a uniform 
phase-coherent reference state with a constant concentration $c_0$ and a 
uniform phase of the oscillators.}
We now construct the generic coupled hydrodynamic equations for the 
two slow 
variables - local fluctuations $\phi({\bf x},t)$ in the phase and concentration 
fluctuations 
$\delta c ({\bf x},t)=c({\bf 
x},t)-c_0$.  We allow for  
advection of $\phi$ and $c$ by 
an incompressible velocity $\bf v$. 
{ Field $\phi$ being the phase fluctuations in a phase-coherent state, is a 
nonconserved broken 
symmetry variable, whereas $\delta c$ is a conserved density. These 
considerations} together with symmetry arguments 
(invariance under translation, rotation 
and a spatially constant shift $\phi\rightarrow \phi + const.$) dictate the 
general forms of the equations  of $\phi$ and $\delta c$. The most 
general coupled dynamical equations for $\phi$ and $\delta c$, where they {\em 
mutually} affect each other, are of the form
\begin{eqnarray}
 &&\frac{\partial\phi}{\partial t}+\frac{\lambda}{2}  
({\boldsymbol\nabla}\phi)^2 +\lambda_1 {\bf v}\cdot {\boldsymbol\nabla}\phi 
=\tilde \Omega (c) +\kappa\nabla^2\phi +\theta,\label{phieq}\\
 &&\frac{\partial \delta c}{\partial t}+\lambda_1 {\bf 
v}\cdot{\boldsymbol\nabla}\delta c= D_c\nabla^2\delta c + \lambda_2 
\nabla^2\phi + {\boldsymbol\nabla}\cdot {\bf f},\label{deneq}
\end{eqnarray}
 in the hydrodynamic limit.
Terms $\lambda({\boldsymbol\nabla}\phi)^2,\tilde\Omega(c)$ in (\ref{phieq}) and 
$\lambda_2\nabla^2\phi$ in (\ref{deneq}) are {\em active terms}, i.e., of 
nonequilibrium origin. These are {\em forbidden in equilibrium} due to the 
invariance 
of an underlying free energy functional $\mathcal F$ under $\phi\rightarrow 
\phi + const.$. 
In the present model, this invariance must be demanded at the 
level of the equations of motion and hence the above active terms are 
permitted in (\ref{phieq}) and (\ref{deneq})~\cite{argument}. {
In Eq.~(\ref{phieq}), we have neglected a subleading cross-coupling term of the 
form $\sim \nabla^2 \delta c$ in the hydrodynamic limit; see below.}
Equations (\ref{phieq}) and (\ref{deneq}) generalize the nonconserved 
relaxational dynamics of the local phase of the nearly phase coherent classical 
XY model; parameter 
$\kappa>0$ is the analog of the spin stiffness of the classical XY 
model~\cite{classicalxy}.
%Here, parameters  $\lambda,\lambda_1,\tilde\Omega$ and $\kappa$ are in general 
%concentration-dependent.
 The $\lambda_1$-terms in (\ref{phieq}) and (\ref{deneq}) above represent 
advection by $\bf v$, and the 
$\lambda$-term in (\ref{phieq}) is a nonequilibrium term  related to the 
well-known 
complex Ginzburg Landau model~\cite{cgle,erwin-cgle} or the dissipative 
Gross-Pitaevskii 
equation for a polariton condensate~\cite{polariton}.
Expanding about $c=c_0$, we 
write 
\begin{equation}
\tilde\Omega (c)= 
\tilde\Omega_0 + \tilde\Omega_1\delta c+\tilde\Omega_2(\delta c)^2 + 
\tilde\Omega_3 (\delta c)^3,\label{omegaeq}
\end{equation}
neglecting other higher order terms.  
Parameters 
$\tilde\Omega_1,\tilde\Omega_2,\tilde\Omega_3$ can be positive or 
negative; without any loss of generality we set $\tilde\Omega_1>0$. 
Parameters $\lambda,\lambda_1$ and $\kappa$, in general functions of $c$, 
upon 
expanding 
about $c=c_0$, yield additional nonlinear terms that are subleading in 
a 
scaling sense (i.e., leave the scaling properties unaffected). Hence, 
we 
ignore their $c$-dependences.  In the limit of 
spatially constant $c$, $\tilde\Omega 
(c)$ may be absorbed by a frequency shift, that 
yields, for ${\bf v}=0$, the 
Kardar-Parisi-Zhang (KPZ) equation for $\phi$~\cite{kpz}. 
Additionally, if $\lambda=0$, Eq.~(\ref{phieq}) 
reduces to the standard relaxational
equation of motion for 
the phase in the classical XY model~\cite{classicalxy}. For 
$\lambda_2=0$, $\delta c$ follows a  
diffusion-advection equation, 
independent
of $\phi$; with $\lambda_2\neq 0$, the  particle mobility is deemed {\em 
active}. Equation (\ref{deneq}) 
implies a concentration current 
\begin{equation}
 {\bf J}_c=- \left[D_c{\boldsymbol\nabla}\delta c -\lambda_1 {\bf v} \delta c+ 
\lambda_2 
{\boldsymbol\nabla}\phi + {\bf f} \right]. \label{curr}
\end{equation}
 Noises $\theta$ and $\bf f$~\cite{noise2} are zero-mean 
Gaussian-distributed with variances $\langle\theta({\bf 
x},t)\theta(0,0)\rangle=2D\delta({\bf x})\delta(t)$ and $\langle f_\alpha ({\bf 
x},t) f_\beta (0,0)\rangle = 2D_1 \delta_{\alpha\beta} 
\delta ({\bf x})\delta (t)$, respectively;  in a nonequilibrium situation, 
$D,\,D_1$ have dimensions of  temperature $T$ and are in general unequal. 
We have ignored cross-coupling terms of purely equilibrium origin in 
Eqs.~(\ref{phieq}) and (\ref{deneq}) above as they are irrelevant (in a scaling 
sense) to the active terms in (\ref{phieq}) and (\ref{deneq}) in the long 
wavelength limit. Equations (\ref{phieq}) and (\ref{deneq}) generalize the 
relaxational dynamics of the classical XY model to any active system having XY 
symmetry with mobility.
 
For a frictional flow, $\bf v$ follows generalized Darcy's 
law~\cite{darcy}, { that here includes the leading order 
symmetry-permitted 
feedback of $\phi$ on $\bf v$~\cite{o_n,modelH}},
%\begin{equation}\label{veq}
% \eta\nabla^2 v_\alpha -\zeta v_\alpha = \nabla_\alpha \Pi + 
%\alpha_0\frac{\partial} {\partial x_\gamma} (\frac{\partial\phi}{\partial 
%x_\gamma} \frac{\partial\phi}{\partial x_\alpha}) + g_\alpha.
%\end{equation}
%Here, $\eta$ is the fluid viscosity, $\zeta$ a friction coefficient, $\Pi$ 
%the generalized pressure and $\alpha_0$ is a coupling constant which may be 
%positive or negative. Noise $g_\alpha$ is a Gaussian white noise with zero 
%mean 
%and a variance $\langle g_\alpha ({\bf x},t) g_\beta(0,0)\rangle = 2D_2 
%P_{\alpha \beta}
%(-\nabla^2)\delta ({\bf x})\delta (t)$. Imposing incompressibility 
%${\boldsymbol\nabla}\cdot {\bf v}=0$, we obtain
\begin{equation}
-\zeta v_\alpha= \alpha_0 P_{\alpha\beta} \frac{\partial} 
{\partial x_\gamma} (\frac{\partial\phi}{\partial 
x_\gamma} \frac{\partial\phi}{\partial x_\beta}) + 
P_{\alpha\beta}g_\beta.\label{genstokes}
\end{equation}
Here, $P_{\alpha\beta}=\delta_{\alpha\beta} -\frac{\nabla_\alpha 
\nabla_\beta}{\nabla^2}$ is  the transverse projection operator and $\zeta$ is a 
friction coefficient~\cite{addicomment}. 
 For a nonequilibrium model, coupling 
$\alpha_0$ has no restrictions on its sign~\cite{o_n}. Noise $g_\alpha$ is a 
zero-mean, Gaussian 
white noise
with a variance $\langle g_\alpha ({\bf x},t) g_\beta(0,0)\rangle = 
2 D_3\zeta\delta_{\alpha\beta}
\delta ({\bf x})\delta (t)$. $D_3$ again has the dimension of temperature.

% We discuss both 
%the cases separately. % the former case, $\lambda_1$ in (\ref{phieq}) and 
%(\ref{deneq}) are irrelevant in a scaling sense; in the second case, it 
%remains 
%relevant and $\bf v$ may be eliminated by using (\ref{genstokes}) to obtain 
%two coupled equations for $\phi$ and $\delta c$ only.

%Active terms  $\tilde \Omega (c)$ and $\lambda_2\nabla^2\phi$ in 
\subsection{The active terms}

 We now discuss the origin and physics of the active terms in more 
details. In Eq.~(\ref{phieq}), if we ignore the time-dependence of $\delta c$, 
the function $\tilde\Omega (c)$ becomes the natural frequency of the 
oscillator. This, in a discrete lattice-gas representation, implies that the 
natural frequency of a particular oscillator is nonuniform and a {\em local 
property}, 
i.e., it depends upon the number of the oscillators in its neighborhood. This 
is a generalization of the well-known Kuramoto model for identical phase 
oscillators (i.e., with the same natural frequency)~\cite{kuramoto1,kuramoto2}.
Depending upon the function $\tilde\Omega (c)$, an oscillator either rotates 
faster or slower as the number of oscillators in its neighborhood changes. 
Consider now the other active term $\lambda_2\nabla^2\phi$ in (\ref{deneq}). 
This corresponds to a current contribution $-\lambda_2{\boldsymbol\nabla}\phi$ 
in ${\bf J}_c$. Thus, neighboring oscillators will move towards or go away from 
each other if there is a phase difference between them, constituting an active, 
$\phi$-dependent current with a magnitude set by $\lambda_2$. Depending on the 
sign of $\lambda_2$, this active current either reinforces or goes against the 
usual diffusive current $D_c{\boldsymbol\nabla} \delta c$.

In the equilibrium limit, the system may be described by a free energy 
$\mathcal F$ given by
\begin{equation}
 {\mathcal F} = \int d^2x [\frac{\kappa}{2}({\boldsymbol\nabla}\phi)^2 + 
A\delta c \nabla^2\phi + \frac{B}{2}(\delta c)^2].\label{free}
\end{equation}
 Here, $A$ and $B$ are thermodynamic coefficients. The sign 
of $A$ is arbitrary, while $B$ is always positive.
Free energy (\ref{free}) yields (assuming simple relaxational dynamics, 
ignoring any advection for simplicity)
\begin{equation}
 \frac{\partial\phi}{\partial t}=-\frac{\delta {\mathcal F}}{\partial {\mathcal 
\phi}} 
+\theta = -[-\kappa\nabla^2\phi + A\nabla^2\delta c] +\theta,\label{eq1}
\end{equation}
and
\begin{equation}
 \frac{\partial \delta c}{\partial t} = 
\nabla^2 \frac{\delta {\mathcal F}}{\delta (\delta c)} + \boldsymbol\nabla 
\cdot{\bf f}= B\nabla^2 \delta c + A\nabla^4\phi + 
{\boldsymbol\nabla}\cdot{\bf f}.\label{eq2}
\end{equation}
The linear cross terms are clearly subleading to the active terms 
$\tilde\Omega_1(c)$ and $\lambda_2\nabla^2\phi$ in Eqs.~(\ref{phieq}) and 
(\ref{deneq}) above. In fact, if we insist on generating these active 
terms from $\mathcal F$, 
we may consider adding terms, e.g., of the form $\tilde\Omega_1 (c)\phi$ in 
$\mathcal 
F$, that generates a term $\tilde \Omega (c)$ in Eq.~(\ref{phieq}) of the main 
text, 
but manifestly breaks the invariance under $\phi\rightarrow \phi + 
const.$, which is not acceptable. This establishes the active origin of the 
term $\tilde\Omega_1(c)$ and similarly of $\lambda_2\nabla^2\phi$ in 
Eqs.~(\ref{phieq}) 
and (\ref{deneq}), respectively, above.

 Equations (\ref{phieq}) and (\ref{deneq}) serve as good 
representations for different real systems.  Consider a 2d 
array of identical water droplets, which contain the reactants of an 
oscillatory chemical (e.g., Belousov-Zhabotinsky) reaction, separated by oil 
gaps; see, e.g., Ref.~\cite{toiya1}, with $\phi$ and $c$, respectively, being 
the phase of the oscillatory reaction and catalyst concentration.  Or
consider a layer of oriented chiral live
cytoskeletal acto-myosin extract resting on a solid substrate. For fully 
oriented actin filaments (in the 
limit of large 
Frank's constant~\cite{jacques-book}, or for length scales smaller than the 
threshold of spontaneous flow instabilities~\cite{rafael}), polarity 
fluctuations may be neglected, and $\phi$, that 
describes chirality of 
actin and $c$, the concentration of actin filaments are the slow variables. 
In 
yet another general
biological motivation of our theory, the vertebrate segmentation 
clocks, $\phi$ 
and $\delta c$ represent, respectively, the local phases of genetic 
oscillations 
and the concentration of the migrating cells or the signaling 
molecules~\cite{vertebrate}. 
In all these 
examples, active terms $\tilde\Omega (c)$ in (\ref{phieq}) and 
$\lambda_2{\boldsymbol\nabla\phi}$ in current ${\bf J}_c$ model generic active 
interplay between phase and concentration fluctuations. 
%(\ref{curr}). %In an equivalent discrete lattice-gas 
%representation of this model comprising of suitably discretized $\phi$ and $c$ 
%on a 2D lattice, the $\lambda_2$-term models 2D generalized symmetric 
%exclusion processes~\cite{sep} that depends upon the local variations in 
%$\phi$.
 %In chiral actin layer or oscillating chemical reactions, 
 In general, all the active coefficients
$\lambda,\tilde\Omega_0,\tilde\Omega_1,\tilde\Omega_2,\tilde\Omega_3,\lambda_2$ 
should depend on   $c_0$, the mean concentration of the diffusing active 
particles.  %In vertebrate 
%segmentation clocks, these couplings should also depend upon delta-notch 
%signaling~\cite{delta}.
 Diffusivity $D_c$ should contain both thermal (equilibrium) and active 
contributions; see, e.g., 
Ref.~\cite{abhik-diff}. 

{ We now compare our model equations with those 
that describe active fluid with orientational degrees of freedom, {\em viz.}, 
the equations of the local polar
order parameter or orientation field $\bf p$ and the concentration of the 
active particles~\cite{active-fluid,john}. While both the systems are 
concerned with the question of order in 2d, there are notable differences 
between 
the two. Our model equation (\ref{phieq}) that generalizes the Kuramoto model 
equation, necessarily applies to phases of oscillators or rotors, i.e., to 
microscopic 
oscillatory
degrees of freedom. Such a collection of oscillators has no notion of local 
orientation or polarity in the physical space. This is  
quite different 
from the 
 active fluids~\cite{active-fluid,john}, where the local polarity 
 describes local orientation of the 
underlying polar or nematic degrees of freedom (i.e., actin filaments, birds 
or fishes).  Furthermore, polar ordered 
active fluids are generically characterized by 
systematic macroscopic motion along the direction of order, where as the 
oscillators considered here are diffusively moving, devoid of any systematic 
large-scale movement. The one particular
case of active fluid models where our model should be relevant is {\em chiral 
active fluids}, where the actin filaments have chirality given by a phase 
variable~\cite{sebastian}. A fully orientationally ordered chiral active 
fluid with very large Frank's constants (that suppress any orientational 
fluctuations) without any large scale motion should be described only by the 
phase
and the local concentration of the active particles. For such a system
our model equations should form a valid 
description.
 Lastly, the ordered state of an active fluid is necessarily
anisotropic due to the macroscopic preferred orientation across the 
system. In 
contrast, a globally synchronized state of phase oscillators like ours is 
perfectly 
isotropic in the physical space. }

%{\em Results:} 
%It is instructive to extract the various time-scales from  
%(\ref{phieq}) and (\ref{deneq}) after linearizing about $c=c_0$. 
\section{Results}\label{result}
\subsection{Fast switching regime}
\subsubsection{Linear theory}

We now analyze Eqs.~(\ref{phieq}) and (\ref{deneq}) to ascertain 
the degree of global synchronization in the model. It is illuminating to first 
consider the linearized version of Eqs.~(\ref{phieq}) and (\ref{deneq}). 
We linearize Eq.~(\ref{phieq}) about $c=c_0$, and  define 
time 
scales $\tau_\phi (q)=1/(\kappa q^2)$ and $\tau_c(q)=1/(D_cq^2)$, where 
$\bf q$ 
is a Fourier 
wavevector; thus $\tau_\phi$ is the time-scale of isolated phase 
fluctuations (i.e., in the absence of any coupling with $c$), where as $\tau_c$ 
is the time-scale in which isolated particles diffuse, or the network evolves. 
Now eliminate $\delta c$ in (\ref{phieq}) to obtain (in the 
Fourier space)
\begin{equation}
 -i\omega \phi = -\kappa q^2 \phi -\frac{\lambda_2\tilde\Omega_1 q^2}{-i\omega 
+ D_c q^2} \phi +\frac{i\tilde\Omega_1 {\bf q\cdot f}}{-i\omega + D_c q^2} + 
\theta. \label{eqlinphi}
\end{equation}
 Here, $\omega$ is the Fourier frequency. The two time-scales $\tau_\phi$ and 
$\tau_c$ can compete with each 
other with 
two asymptotic limits $\tau_\phi \gg\tau_c$ (fast switching regime in the 
network language~\cite{pagona}) and $\tau_\phi \ll \tau_c$ (slow switching 
regime). 
We note 
that 
recent studies on synchronization in 1d using agent based 
models~\cite{peruani} indicate that large oscillator diffusivities tend to
enhance the degree of global synchronization. A large $D_c$ implies a small 
$\tau_c(q)$ for fixed $q$.  
Taking cue from this and in order to extract
the activity-dependence of synchronization in the most dramatic way, we 
consider the limit $D_c\rightarrow \infty$, or equivalently, $\tau_c\rightarrow 
0$; clearly $\tau_c\ll\tau_\phi$ for a finite $\kappa$. In this 
limit, Eq.~(\ref{eqlinphi}) simplifies to (in the time domain)
\begin{eqnarray}
 \frac{\partial\phi}{\partial t} &=& -\kappa q^2\phi - \frac{\tilde\Omega_1 
\lambda_2}{D_c}\phi + \theta + 
\frac{i{\bf q}\cdot {\bf f} \tilde \Omega_1}{q^2D_c} \nonumber\\
 &=& -\kappa q^2\phi -\gamma \phi +\theta + 
 \frac{i{\bf q}\cdot {\bf f} \tilde \Omega_1}{q^2D_c}, \label{effphi}
\end{eqnarray}
where $\gamma=\tilde\Omega_1\lambda_2/D_c$ is an active coefficient. 
Equation (\ref{effphi}) allows us to extract yet another time-scale 
$\tau_\times = 
\frac{D_c}{\tilde\Omega_1 \lambda_2}=\frac{1}{\gamma}$ ($\gamma$ has the 
dimension of inverse 
time); $\tau_\times$ is infact the time-scale of phase fluctuations due 
to the active coupling of $\phi$ with concentration fluctuation $\delta c$. We 
assume $\gamma\sim O(1)$, i.e., $\tau_c/\tau_\times\rightarrow 0$. 
(Note that for $\gamma$ to dominate the long wavelength dynamics 
 of $\phi$, $|\gamma| > \frac{4\pi^2 \kappa}{L^2}$. 
We present most of our results in this limit.  This is realizable for 
large enough $L$ with sufficiently large $\tilde\Omega_1\lambda_2$.)
Since active coefficients $\tilde\Omega_1$ and $\lambda_2$ are formally 
independent of $D_c$, this can be realized by letting 
$\tilde\Omega_1\lambda_2\rightarrow$ large with a large $D_c$. 
Notice 
that in this 
fast switching regime,  the dynamics of $\delta c$ is effectively slaved to 
$\phi$.
 We now set out to
calculate $\Delta$ below in the limit of fast dynamics of 
$\delta c$.
\par
%We find
%\begin{equation}
 %$\delta c= \frac{-\lambda_2 q^2 \phi + i{\bf q \cdot f}}{ D_c q^2}$,
%\end{equation}
%This together with Eq.~(\ref{phieq}) produces an effective (linear) equation 
%for 
%$\phi$; 
%see SM (Sec.~IV). 
Evidently for a positive $\tilde \Omega_1$,  Eq.~(\ref{effphi}) 
is linearly 
unstable if {\em active damping}
$\gamma<0$ ($|\gamma|$ : growth rate);
i.e., if $\lambda_2<0$; else, 
it is linearly stable ($\gamma$ : decay rate).  For a positive $\tilde 
\Omega_1$, 
thus, 
$\lambda_2 >0 (<0)$ implies that any local excess of diffusive species reduces 
(enhances) any local nonuniformity in $\phi$~\cite{micro1}. 
%{\bf The discussion on linear instabilities seems missing. - etao lekha ache 
%tex 
%file ei. comment out kora ache ektu niche.
%NA BOSS OTA ALADA. AMI ADD KORE DIYECHHI OPOR-E.}
%The effective dynamics of $\phi$ is clearly controlled by two noises, {\em 
%viz.}, $\theta$ 
%and
%$f^{\prime} (=\frac{i \tilde \Omega_1 q \cdot f}{D_c q^2})$; see SM for 
%details. 
\par

Now, consider 
\begin{equation}
\Delta = \int_{1/L}^\Lambda \frac{d^2q}{(2\pi)^2} \langle |\phi 
({\bf 
q},t)|^2\rangle=\int_{1/L}^{\Lambda}\frac{d^2q}{(2\pi)^2}\frac{d\Omega}{2\pi}
\frac{\langle|\theta|^2 + |f'|^2\rangle }{\Omega^2+\gamma^2}
\end{equation}
for $\gamma>0$ 
in the 
linearized theory together with $\tau_c\rightarrow 0$ and  
$\tau_\times\sim O(1),\tau_c/\tau_\times\rightarrow 0$. Here, $f'= 
\frac{\tilde\Omega_1 iq_\alpha f_\alpha}{q^2 D_c}$,  $\phi({\bf 
q},t)$ is the Fourier transform of $\phi ({\bf x},t)$.
 We define a length scale $L_c$ given by the relation 
 \begin{equation}
 \frac{D_1 \tilde 
\Omega_1^2}{D_c^2 D}=1/L_c^2.\label{lc}
\end{equation}
 Since $L_c$ depends explicitly on the active coefficient 
$\tilde\Omega_1$, it can be tuned by the active processes. In particular, $L_c$
can be made very large for small $\tilde\Omega_1$.
The nature of order depends sensitively on 
the dimensionless ratio $L/L_c$, as we establish below.
%{\bf It is the ratio $L/L_c$ and the parameterized length 
%scale $L_c$ that
%determines the nature of order of the oscillators.}

\begin{figure}
\includegraphics[height=6cm]{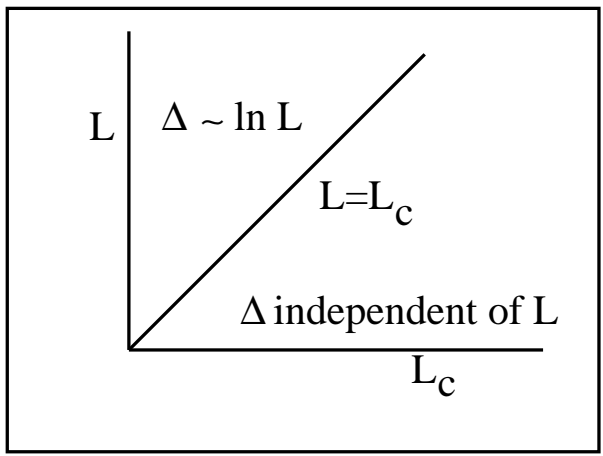}
\caption{ Behavior of $\Delta$ in the $L_c-L$ plane in the linear theory 
(with $\gamma>0$) in the fast switching regime. The straight
line refers to $L=L_c$; see text.}  \label{linstab1}
\end{figure}

%For large $L_c\rightarrow \infty$ with system 
%size $L < 
%L_c$,
%$\theta$ dominates %over 
%$f^{\prime}$ and hence
%the latter can be neglected~\cite{remark_2}.
%Hence,
For $\frac{L}{L_c} \ll 1$, this yields
\begin{equation}
 \Delta =\int_{1/L}^{\Lambda}\frac{d^2q}{(2\pi)^2}\frac{d\omega}{2\pi}
\frac{\langle|\theta|^2\rangle }{\omega^2+\gamma^2} \approx \frac{D\Lambda^2}{2\pi\gamma},
\end{equation}
 for $L \gg 1/\Lambda$.  Notice that this $L$-independence 
of $\Delta$ holds 
even in 1d. In contrast for large $\frac{L}{L_c} \gg 1$, 
 \begin{equation}
  \Delta=\int_{1/L}^{\Lambda}\frac{d^2q}{(2\pi)^2}\frac{d\omega}{2\pi}
\frac{\langle|f'|^2\rangle }{\omega^2+\gamma^2} \approx \frac{D}{2\pi 
L_c^2\gamma}\ln 
L,
 \end{equation}
 for $L \gg 1/\Lambda$.
 Here, $\Lambda$ is an upper 
wavevector cut-off, $\omega$ is a frequency~\cite{remark_2}. See 
Fig.~\ref{linstab1} for a schematic phase diagram 
(with $\gamma >0$) in the $L_c-L$ plane, showing regions corresponding to LRO 
and QLRO, respectively.

\begin{figure}[htb]
\includegraphics[height=6cm,width=8cm]{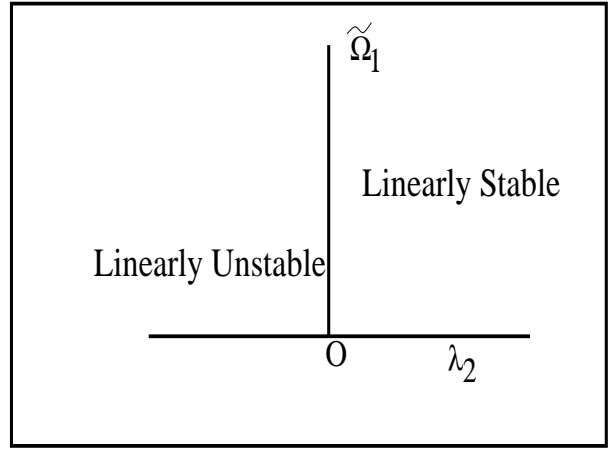}
\caption{Schematic phase diagram in the 
$\lambda_2-\tilde\Omega_1$-plane showing linearly stable and unstable regions 
in the linear theory in the fast switching regime; see text. 
 Symbol $O$ marks the origin $(0,0)$.}\label{linstab}
\end{figure}

 The nature of order can be established from
 the equal-time oscillator 
correlator~\cite{o_n}
\begin{equation}
C_s(r)=\langle\cos [\phi ({\bf x},t)-\phi 
(0,t)]\rangle=\exp[-g(r)]
\end{equation}
for large $r=|\bf x|$ with  
$g(r)=\langle[\phi({\bf 
x},t)-\phi(0,t)]^2\rangle/2$. This yields
\begin{equation}
g(r)=D\Lambda^2/(2\pi\gamma),
\end{equation}
for large $r$ (with $r\lesssim L<L_c$), demonstrating LRO. 
Thus, for a system of size $L\ll L_c$ 
and $\gamma>0$, the system can show LRO in the fast switching regime, such that 
$L\ll L_c$; see Refs.~\cite{frasca} and \cite{peruani} which show numerically 
how increasing mobility can lead to global synchronization. Notice that in our 
model $L_c$ 
may be made arbitrarily large even with a large $D_c$ by adjusting the other 
model parameters; see Eq.~(\ref{lc}) above.
%Thus, for any system with size $L<L_c$ in 1D or 
%2D, the oscillators are globally synchronized~\cite{peruani,frasca}.
 In contrast, for large $r$ with $L_c<r<L$,
 \begin{equation}
g(r)= \frac{D}{2\pi L_c^2\gamma} \ln r,
\end{equation}
hence, showing QLRO 
in the system~\cite{o_n}. 
For $\gamma<0$, only for a finite system size with $L^2 < 
4\pi^2\kappa/|\gamma|$, $\Delta$ is finite, implying SRO~\cite{peruani}. This allows 
us to define a 
persistence length $\xi_1$, given by 
\begin{equation}
\xi_1=2\pi(\kappa/|\gamma|)^{1/2},
\end{equation}
such 
that for $L>\xi_1$, instability ensues.  Thus given the 
activity-dependences of $\gamma$ and $L_c$, the role of active effects in 
facilitating 
or destroying order 
is clearly 
established. 
Evidently, the larger $|\gamma|$ is, the stronger is the linear instability 
($\gamma <0$) or stronger suppression of phase fluctuations (more stable, 
$\gamma >0$); see Fig.~\ref{linstab} for a schematic phase diagram in the $\lambda_2 - \tilde\Omega_1$
plane. Since all of $D_c,\tilde\Omega_1$ and $\lambda_2$ 
are expected to scale with $C_0$, the mean concentration, $\gamma$ should scale 
with $C_0$. Thus any activity-induced synchronization or instability should be 
enhanced in a denser system - a broad feature testable in experiments on 
relevant systems.

%A summary of our results are 
%listed in Table~\ref{table}.} 

\subsubsection{Nonlinear effects}

We now consider the dominant nonlinear effects on the results 
from the linear theory.
 For 2d equilibrium systems with continuous symmetries in their ordered 
phases, this is most conveniently done in terms of a low-$T$ expansion in 2D, 
where a 
dimensionless reduced $T$ (assumed small in the putative ordered phases) plays 
the role of a small parameter~\cite{classicalxy}. This has 
subsequently been 
extended to analogous nonequilibrium systems, see, e.g., Ref.~\cite{o_n}.
Following Ref.~\cite{o_n}, we identify 
$D\Lambda^2/\gamma$ and $D/(L_c^2\gamma)$ 
%and 
%$D_3\lambda_2/(\zeta 
%D_c^2)$ 
as the dimensionless small parameters and restrict 
ourselves up to their linear order expansions, for 
(assumed) small 
fluctuations of $\phi$ and $v_\alpha$ with $\tau_c/\tau_\times\rightarrow 0$.
We replace $\tilde\Omega_3 (\delta c)^3$ in Eq.~(\ref{omegaeq}) by a linear term
$3\tilde\Omega_3 \langle(\delta c)^2\rangle\delta c$ in a Hartree-like 
approximation~\cite{classicalxy} and  substitute it in Eq.~\ref{phieq}. This 
then produces a correction $\overline{\gamma}$ to $\gamma$.
With $\gamma>0$, following
our analysis at the linear order above,
% 
%Hence, %$\tilde \Omega_1$ 
%receives the leading order fluctuation correction from
%$\tilde \Omega_3$. 
 we define an effective  active damping coefficient
\begin{equation}
\gamma_e=\gamma+\overline{\gamma},
\end{equation}
where $ \overline\gamma=\frac{3\tilde \Omega_3 
\langle(\delta 
c)^2\rangle\lambda_2}{D_c}$.
%Now use,
%\begin{equation}\label{c-fluc}
 %\langle(\delta c)^2\rangle=\int \frac{d^d q}{(2\pi)^d} \frac{d\omega}{2\pi} 
%\left[ |A|^2 \langle\theta^2\rangle + |B|^2 \langle|f^{\prime}|^2\rangle 
%\right],
%\end{equation}
%at the lowest order in smallness, where, $A$ and $B$ are complex 
%coefficients~\cite{complex-coeff}. 

Now, in the linear theory  with $\tau_c\rightarrow 0$ and 
$\tau_c/\tau_\times\rightarrow 0$, such that the dynamics of $\delta c$ is 
slaved to that of $\phi$,
\begin{equation}
\delta c=-\frac{\lambda_2}{D_c}\phi + \frac{i{\bf q\cdot f}\tilde\Omega_1}{D_c 
q^2}.\label{deltac1}
\end{equation}
Equation (\ref{deltac1}) may now be used to calculate $\langle|\delta 
c ({\bf q},t)|^2\rangle$ separately for $L<L_c$ and $L>L_c$, since the 
form of $\langle |\phi({\bf q},t)|^2\rangle$ depends on whether $L<L_c$ or 
$L>L_c$.

(i) For $L <L_c$, 
\begin{equation}
\langle\delta c^2\rangle=\frac{\lambda_2^2}{D_c^2} \int_{1/L}^{\Lambda} 
\frac{d^2 q \, d\Omega}{(2\pi)^3} \frac{D}{\gamma} \approx \frac{\lambda_2^2 D 
\Lambda^2}{4\pi\gamma D_c^2}.
\end{equation}

Using this value of $\langle(\delta c)^2\rangle$ the correction to $\gamma$ takes the form 
\begin{equation}
\overline{\gamma}=3\tilde\Omega_3\lambda_2^2 D\Lambda^2/(4 \pi D_c^2 
\tilde \Omega_1),
\end{equation}
which is finite.  For $\tilde\Omega_3>0$, we have 
$\overline{\gamma} >0$, 
thus leaving the results from the linear theory qualitatively unchanged. On the 
other hand,
for $\tilde\Omega_3<0$, this
correction is negative and has the 
potential of introducing instability provided $|\overline{\gamma}|>\gamma$. This 
yields 
a finite instability threshold for $|\tilde\Omega_3|$, given by $\tilde \Omega_{3}^c=\frac{4\pi\gamma\tilde\Omega D_c^2}{3\Lambda^2 \lambda_2^2 D}$.

\begin{figure}[htb]
\includegraphics[width=8.7cm]{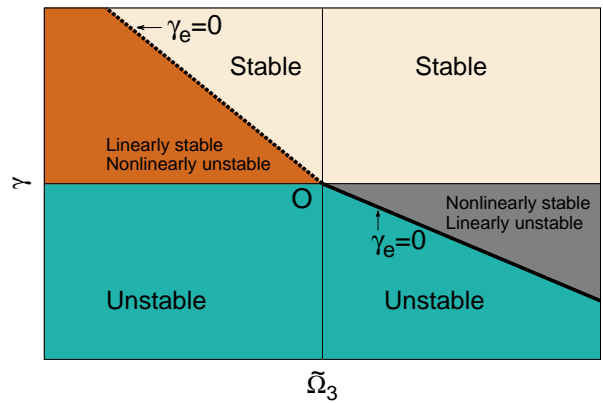}\hfill
\caption{ (Color online) Schematic phase diagram in $\tilde\Omega_3-\gamma$ 
plane ($\zeta> 0$). Symbol $O$ refers to the 
origin $(0,0)$. Regions (i) 
linearly and nonlinearly stable (both $\gamma,\gamma_e>0$), (ii) 
linearly stable, nonlinearly unstable ($\gamma>0,\gamma_e<0$), (iii) 
linearly unstable, nonlinearly stable ($\gamma<0,\gamma_e >0$) and 
(iv) 
linearly and nonlinearly unstable (both $\gamma,\gamma_e <0$) are 
marked (see text). The inclined lines, drawn schematically, are to 
be obtained from the conditions $\gamma_e=0$.}  \label{gamma-phase}
\end{figure}

(ii) For $L \gg L_c$,
$\overline{\gamma} \approx \Gamma \ln L$, diverging logarithmically with $L$; 
\begin{equation}
\Gamma=3\tilde\Omega_3 \lambda_2^2\tilde\Omega_1 D_1/(2\pi 
D_c^4),
\end{equation}
see Appendix.
Thus, $\overline{\gamma}$ necessarily dominates over $\gamma$ for 
sufficiently large $L$; thence, for 
$\tilde\Omega_3<0$ and $\gamma >0$ the instability necessarily sets in for 
a sufficiently large $L$, without any finite threshold for 
$\tilde\Omega_3$.  This nonlinearity induced instability allows us to 
introduce a modified persistence length 
\begin{equation}
\tilde 
\xi_1= 2\pi (\kappa/|\gamma_e|)^{1/2},
\end{equation}
now controlled by  $\tilde\Omega_3$ 
for a fixed $\gamma>0$. Similar
to the calculations for the linear theory, now
with $\gamma_e>0$, we find for 
large $r\lesssim L<L_c$, $g(r)=D\Lambda^2/(2\pi\gamma_e)$. This
confirms
LRO as in the 
linear theory (see above). On the other hand, for 
large 
$r > L_c$, $g(r)=\overline A\ln\ln r$, 
showing that $C_s(r)$ decreases as $1/(\ln r)^{\overline A}$; $\overline A= 
D_1\tilde\Omega_1^2/(\Gamma D_c^2)$. Thus in this case, QLRO 
in the linear 
theory 
gets modified to NLO by the nonlinear effects with a 
spatial decay 
slower than the algebraic decay in QLRO~\cite{o_n}. 
For $\gamma <0$ ($\lambda_2 <0$), linear instability ensues. 
However, nonlinear effects can 
stabilize and suppress this linear instability, provided $\tilde\Omega_3>0$ and 
$\overline{\gamma} > |\gamma|$. Therefore, depending on the signs of 
$\lambda_2$ and $\tilde\Omega_3$, four distinct possibilities emerge, as shown 
in Fig.~\ref{gamma-phase} schematically in the 
$\tilde\Omega_3-\gamma$ plane. 

% We now consider the advective nonlinearities, i.e., the $\lambda_1$-terms 
%in 
The advective nonlinearities in Eqs.~(\ref{phieq}) and (\ref{deneq}) 
generate additional corrections to the model parameters in Eqs.~(\ref{phieq}) 
and (\ref{deneq}).
 This is acheived by eliminating $\bf v$ in Eq.~(\ref{phieq}) with the help of 
Eq.~(\ref{genstokes}), which 
generates finite corrections to $\kappa$ and $\gamma$ for both $L<L_c$ and 
$L>L_c$. See Appendix for some calculational details regarding advective 
nonlinearities.
\par
\begin{figure}[htb]
\includegraphics[width=8.5cm]{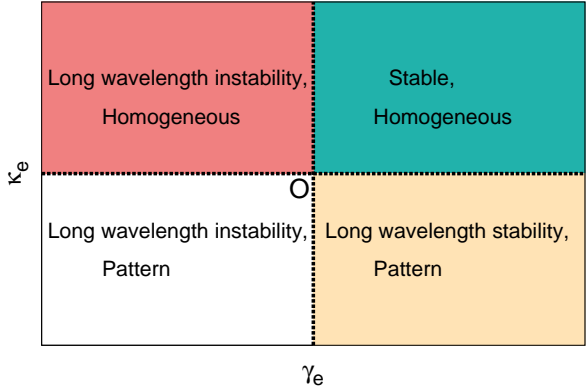}\hfill
\caption{(Color online) Schematic phase 
diagram in $\gamma_e-\kappa_e$ plane. 
Symbol
$O$ marks the origin $(0,0)$. Four distinct macroscopic behaviors are possible 
: (i) 
$\gamma_e,\kappa_e >0$ (Stable, homogeneous), 
$\gamma_e>0,\kappa_e <0$ (long wavelength stability 
with patterns at $O(q^2)$, (iii) $\gamma_e<0,\kappa_e>0$ (long 
wavelength unstable with no 
pattern) and (iv) $\gamma_e,\kappa_e<0$ (long wavelength instability with patterns 
at $O(q^2)$); 
see text. }  \label{phasenofriction}
\end{figure}

 
For $\lambda_1\alpha_0$ sufficiently negative, $\gamma_e$ and 
$\kappa_e$ that now include additional corrections from $\lambda_1$, 
may become negative and 
thus lead to long wavelength instabilities and pattern formations. 
 For instance, for $\gamma_e>0$ and $\kappa_e<0$, finite wavevector 
instabilities are produced in the system at $O(q^2)$, while 
 remaining stable at
$O(q^0)$. The 
instability at $O(q^2)$ for $\kappa_e<0$ should be suppressed at very 
high $q$ by a 
stabilizing generic fourth order spatial derivative term $-\kappa_4 
\nabla^4\phi (\kappa_4>0)$ in the rhs of Eq.~(\ref{phieq}) (here neglected; 
see Ref.~\cite{lauter}). For 
$\lambda_1\alpha_0>0$, these corrections do not affect the 
scaling of $\Delta$.
%; $L$ being the linear length scale associated with the 
%system.
%\begin{figure}[htb]
% \includegraphics[height=4cm,width=4cm]{model.eps}\hfill  
%\includegraphics[width=8.0cm]{delta-L.eps}\hfill
%\includegraphics[width=4.2cm]{L-Lc.eps}
%\caption{(Color online) Variation of $\Delta$ with $L$ ($\zeta>0$). Shaded 
%region 
%corresponds to LRO to NLO crossover (see text).{\bf DROP THIS FIG}}  
%\label{delta-L}
%\end{figure}
%In the above, we have assumed  a 
%frictional background. 
%%We now consider non-linear stability of an unstable system ($\gamma_e<0$).%%
 The  
regions in phase space where 
$\kappa_e$ is negative should display patterns in the steady state. Our 
hydrodynamic theory, based on retaining only the lowest order gradients and low 
order nonlinear terms cannot determine the steady state patterns. The detailed 
nature of the patterns 
should depend upon the higher order terms neglected here; see, e.g., 
Refs.~\cite{lauter,patterns} for a
related recent study. How the steady state patterns depend upon the feedback of 
the phase fluctuations on mobility remains an important question to be studied 
in the future.
The four possible macroscopic behavior  in the $\gamma_e-\kappa_e$ 
plane are shown
schematically in Fig.~\ref{phasenofriction}: (i) long wavelength stable, homogeneous (green) with 
$\gamma_e>0,\kappa_e>0$, (ii) long wavelength instability but homogeneous 
(red) with $\gamma_e<0,\kappa_e>0$, (iii) long wavelength instability with 
patterns (white) with $\gamma_e<0,\kappa_e<0$, and (iv) long wavelength stability 
with patterns (yellow) with $\gamma_e>0,\kappa_e<0$.

%\begin{figure}[htb]
% \includegraphics[height=4cm,width=4cm]{model.eps}\hfill  
%\includegraphics[width=7.5cm]{g-q.eps}\hfill
%\caption{(Color online) A plot of $g(q)$ vs $q$. In a unit of 
%time given by $|\gamma|=1$ and nondimensionalizing $q$ by $L$, curves marked
%C ($\gamma_e=-0.6, \kappa_e=-1.8$) (blue) and D ($\gamma_e=0.6, \kappa_e=-1.8$) 
%(pink) represent unstable and stable systems with patterns at $O(q^2)$,
%respectively, for small $q$. A ($\gamma_e=0.6, \kappa_e=1.8$) (red) and B 
%($\gamma_e=-0.6, \kappa_e=1.8$) (green) represent stable and unstable 
%homogeneous systems,
%respectively, for small $q$; see text. $\kappa_4=0.5$ in each case. The 
%horizontal solid (black) line 
%represents $g(q)=0$.
%The relevant length scales of $g(q)$ are determined by 
%$\gamma_e,\kappa_e$ and $\kappa_4(>0)$.}  \label{pattern}
%\end{figure}

%\begin{figure}[htb]
%\includegraphics[width=6cm]{kapp-gamm_ph.eps}\hfill
%\caption{}  \label{pattern}
%\end{figure}

%It is evident that $\frac{\tilde \Omega_1 \lambda_1 \alpha_0}{\nabla^2} \nabla_\alpha \left[\frac{P_{\alpha \beta}}{\eta \nabla^2}\nabla_\gamma(\nabla_\beta \phi \nabla_\gamma \phi) \frac{\lambda_2 \phi}{D_c}\right]$
%produces a correction to $\kappa$ with a factor of $\frac{\tilde \Omega_1 \lambda_1 \alpha_0}{q^2}$; equivalently, it corrects the bare damping, $\gamma$.

\par
%schematic phase diags?
\subsection{Slow switching regime}

\begin{figure}[htb]
\includegraphics[width=8cm]{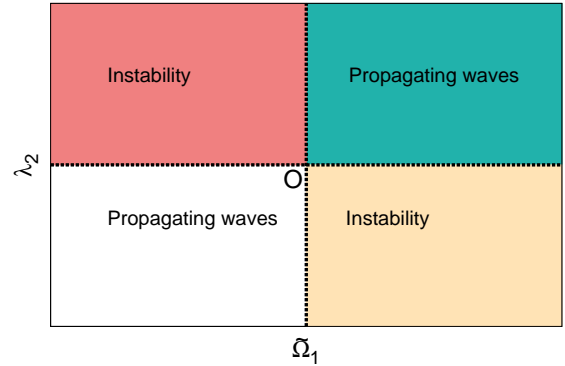}\hfill
\caption{(Color online) Schematic phase 
diagram in $(\tilde\Omega_1,\lambda_2)$ plane in the linearized theory with 
$D_c\rightarrow 0$. 
Symbol
$O$ marks the origin $(0,0)$. Two distinct macroscopic behaviors are possible 
: (i) 
$\tilde\Omega_1,\lambda_2>0,\tilde\Omega_1,\lambda_2 <0$ : Propagating waves, 
 and (ii) $\tilde\Omega_1>0,\lambda_2<0$ and $\tilde\Omega_1<0,\lambda_2>0$ : Instability in the system.
see text. For the convenience of the reader, negative values of 
$\tilde\Omega_1$ have also been shown.}  \label{smallDc}
\end{figure}

We now briefly discuss the limit of slow switching regime, 
$\tau_c\gg\tau_\phi$. In order to extract the physics in this regime most 
effectively, we consider the limiting case and set $D_c=0$, 
 in Eq.~(\ref{deneq}), so that 
$\tau_c\rightarrow \infty,\,\tau_\times \rightarrow 0$. 
Notice that in our model, even with $D_c=0$, the concentration dynamics 
does not freeze; this is essentially due to its coupling with the gradient of 
$\phi$ via the $\lambda_2$-term in Eq.~(\ref{deneq}). Then, assuming
time dependences for $\phi$ and $\delta c\sim\exp (\tilde\Lambda t)$,
we find
\begin{equation}
 \tilde\Lambda \approx \frac{1}{2}\left[-q^2 \kappa  \pm 2 
q\sqrt{-\tilde\Omega_1 
\lambda_2} \right]
\end{equation}
Hence, depending upon the sign of $\lambda_2$ (with $\tilde\Omega_1>0$), the 
system may exhibit 
either
damping along with underdamped waves ($\lambda_2>0$) with speed 
$v_0=\sqrt{\tilde\Omega_1\lambda_2}$,
or long-wavelength instability ($\lambda_2<0$)~\cite{peruani,remark_1}. 
In the linearly stable region, we find in the linearized theory, $\Delta\sim 
(\tilde\Omega_1^2 D_1 + Dv_0^2)(\ln L)/\kappa$, yielding QLRO. This 
corresponds to $C_s(r)$ decreasing as $1/r^\psi,\,\psi=(\tilde\Omega_1^2 D_1 
+ D\tilde\Omega_1\lambda_2)/\kappa$. This is 
unaffected by the leading order nonlinearities. For $\lambda_2 <0$, similar to 
the discussions
above, a persistence length $\xi_2$ may be defined as 
$\xi_2=\kappa/\sqrt{\tilde\Omega_1|\lambda_2|}$ such that for 
system size $L>\xi_2$, $\Delta$ diverges, implying SRO. 
Lastly, coupling 
$\lambda$ remains irrelevant throughout (in a scaling sense) in all the cases 
discussed above.
A phase diagram in the $\tilde\Omega_1-\lambda_2$ plane is given in 
Fig.~\ref{smallDc}, showing the linearly stable and unstable regions.

%{\bf small D limit, pattern formations in the nonlinearly unstable cases}

\subsection{System behavior with autonomous mobility}

Our results for both the fast and slow switching regimes 
crucially depend on the coupling $\lambda_2$; indeed, $\gamma=0$ if 
$\lambda_2=0$. In the 
agent-based models of Ref.~\cite{frasca,peruani,pagona}, agent 
mobility is fully {\em autonomous}, 
unaffected by the local phase, and hence these models imply $\lambda_2=0$ in 
the continuum. Nevertheless, even for those 
models it has been observed that for a sufficiently large $D_c$, oscillators 
tend to synchronize in systems of finite size. We now discuss this effect 
heuristically in a coarse-grained description. In these agent based models, 
while mobility is autonomous, 
phase fluctuations are still affected by diffusion. This happens essentially 
due to the particle diffusion enabling a test particle to have a larger number 
of contacts in its locality than without diffusion. This should {\em 
renormalize} the elastic modulus $\kappa$ that should now become a local 
quantity $\kappa(c)$ depending upon the local concentration fluctuation $\delta 
c$. Thus, in the coarse-grained hydrodynamic limit, expanding for small $\delta 
c$, we can phenomenologically replace $\kappa$ in 
Eq.~(\ref{phieq}) by
\begin{equation}
 \kappa(c)=\kappa_0 + \tilde g\delta c +g (\delta c)^2,
\end{equation}
where $\tilde g$ and $g$ are coupling constants. Averaging over $\delta 
c$-fluctuations yields an effecting spin stiffness $\kappa_e$ given by
\begin{equation}
 \kappa_e=\kappa_0 +g\langle (\delta c)^2\rangle.
\end{equation}
For a given short range interaction, in a coarse-grained picture coupling $g$ 
should scale with the number of 
interactions $n_0$ in a unit time. Since $n_0$ should rise  with $D_c$, $g$ 
should also rise with $D_c$. Thus, for a large enough $D_c$, $\kappa_e\gg (\ll)
\kappa_0$ for $g > (<)0$. Clearly, with $g>0$ $\phi$-fluctuations are 
significantly suppressed for a large enough $D_c$, such that in a finite system 
the oscillators appear synchronized.

%{\em Conclusions:}
%We have, thus, shown within a generic hydrodynamic theory for a 
%collection of identical oscillators that the 
%interplay between  
%activity and the diffusivity of a dynamically coupled active species controls 
%the degree of order and 
%synchronization or 
%desynchronization of the oscillator phases at 2D. 

\section{Summary and outlook}\label{summ}

 To summarize, we have constructed the dynamical equations for active 
hydrodynamics of a collection of nearly phase-coherent diffusively moving 
oscillators. Depending 
upon the details of the active processes, the system can display a wide-ranging
nature of order, e.g., LRO, NLO and SRO. In particular, a large mobility can 
facilitate LRO for appropriate choices of the other model parameters. These are 
robust features, unaffected 
by an advecting velocity field. Our theory is generic, and 
is applicable to any nearly phase-ordered system of diffusively moving 
oscillators. We generally
predict that the 
degree of synchronization and the nature of order in a 2d
 collection of mobile oscillators can be controlled by
the underlying microscopic active 
processes. In the equilibrium limit, a collection of 
oscillators can only display QLRO at low $T$~\cite{classicalxy}, and hence can 
be only partially synchronized. Thus  
LRO, NLO and SRO in this model are entirely of active origin. For 
an isolated, free standing system, $-\zeta v_\alpha$ in 
Eq.~(\ref{genstokes}) should be replaced by  $\eta\nabla^2 
v_\alpha$, $\eta$ being the fluid viscosity. The system does not show any
new qualitative form of order in 
this limit.  Oscillatory 
chemical 
reactions, oriented chiral live cytoskeletal extracts, {\em in-vivo} 
vertebrate 
segmentation clocks can thus be 
in a phase coherent or decoherent state, 
depending 
upon the 
details of the underlying active processes; the feedback of phase 
fluctuations on the mobility is expected to play an important role in the 
ensuing large scale behavior. Independent of the precise numerical 
values 
of model parameters,
the general structure of the phase diagrams are robustly testable in related 
chemical and biologically inspired systems~\cite{toiya1,toiya2,frank-exp}. 
In experiments on {\em in-vitro} chiral cytoskeletal suspensions, $D_c$ may be 
controlled by changing
the viscosity of the solvent; active parameters
$\tilde\Omega_1,\lambda_2,\tilde\Omega_3,\lambda_1$ may be controlled by 
changing $c_0$; the sign of $\gamma$ may be 
controlled by appropriate choices of contractile or extensile 
activties~\cite{active-fluid}. Numerical simulations of models for vertebrate 
segmentation clocks can be used to verify our 
results~\cite{num-vertebrate}. 
% We expect that our theory should lead to new physical understanding about 
%controlling sychronization in a collection of interacting oscillators.
  %{ How couplings with active catalysis affect the order-disorder transition
%for $D>2$, can be studied in the framework of CGLE, coupled with $\delta c$.
%The results should be of interest in bulk systems.}
%We expect our theory to introduce new directions in the physical 
%understanding of the behavior of chemical oscillators in the presence of 
%catalysts. 
%Experimental detection of NLO is expected to be a challenging task, 
%owing to the very weak $L$-dependence of $\Delta$. %{ Nonetheless, in
%experiments, a non-zero 
%$\lambda_2$ may be detected by measuring $\delta c$ correlations and/or, more 
%importantly, cross correlation function  $\langle \phi \delta c\rangle$, that 
%vanishes identically for $\lambda_2=0$, providing stringent tests of our 
%predictions. }  
%The nonequilibrium effects in our model should be controlled by the ATP or 
%catalysts concentrations in cell cyotoskeletal extracts and 
%oscillating chemical reactions, respectively.
%Considering
%the complex 
%dependences of synchronization and order on these active 
%parameters, experimental characterization of relevant 
%physical systems would be challenging.  
Our theory may be 
extended to account for superdiffusion~\cite{markus} with L\'evy 
noises~\cite{levy}. Additional features, e.g., coupling delays and phase 
shifts~\cite{frank2}, 
may be easily
incorporated in our model.
Large-scale
numerical studies will ultimately be needed
to go beyond the perturbative analysis presented
here, and to study the role of topological defects that are neglected above. 
 Our work should inspire new studies on agent-based models of 
synchronization in dynamical networks that would generalize the existing 
studies with autonomous particle 
mobility~\cite{frasca,pagona,peruani,net1,affect}. 
 We look 
forward 
to future
attempts to verify our results in 
controlled experimental set ups. It would also be of interest 
to include large-scale systematic motion of the oscillators, controlled by 
external drive, 
in our model, and investigate how that may affect the nature of ordering 
elucidated here.  We have effectively considered point particles 
without inter-particle interactions or any excluded volume interactions. 
Generalization of our model to groups of oscillators with distinct 
chiralities remains an interesting issue. Our 
model may be extended to include these features in straight forward ways. 
The insight gathered in this work should help design specific 
synchronization strategies in {\em in-vivo} systems with artificial microscopic 
agents. We expect our work to be a significant stepping stone
in developing a general theory for active hydrodynamics of synchronization 
phenomena.

\section{Acknowledgement}

The authors thank 
the Alexander von Humboldt Stiftung, Germany for partial financial support 
through the Research Group Linkage Programme (2016).

\begin{appendix}

\section{Connection with CGLE}
The complex Ginzburg Landau equation (CGLE) describes the coupled dynamics of 
the amplitude and phase of a complex field $Z=Z_0({\bf x},t)\exp [i\phi ({\bf 
x},t)]$, where the amplitude $Z_0$ and phase $\phi$ are real functions of $\bf 
x$ and $t$. 
\begin{equation}
 \partial_t Z=-\frac{\delta {\mathcal F}}{\delta Z^*} 
-i\Gamma_I\frac{\delta 
{\mathcal F}}{\delta Z^*} +\Theta,\label{cgle1}
\end{equation}
where $\Gamma_I$ is real, 
\begin{equation}
 {\mathcal F}=\int d^dx [\frac{\tilde r}{2}|Z|^2 +\frac{g}{2}|{\boldsymbol 
\nabla}Z|^2 
+ u|Z|^4],
\end{equation}
$\tilde r=0$ gives the mean field second order transition temperature, $u>0$.
Here, $\Theta$ is a complex Gaussian noise with zero mean and a variance
\begin{equation}
 \langle\Theta({\bf x},t)\Theta^*(0,0)\rangle = 2D_\xi \delta ({\bf x})\delta 
(t),\,
 \langle\Theta({\bf x},t)\Theta(0,0)\rangle=0.
\end{equation}
If we now set $Z_0=1$, i.e., fixed amplitude for the complex field, $\phi$ then 
follows the real equation
\begin{equation}
 \frac{\partial\phi}{\partial t} = g\nabla^2\phi +\Gamma_I[-\tilde r-2u + 
g({\boldsymbol\nabla}\phi)^2] 
+ \tilde\theta.
\end{equation}
Thus,  Eq.~(\ref{phieq}) may be obtained by considering $\Gamma_I$ as a function 
of 
$c$ and with the identification
$\kappa=g,\tilde\Omega_0=-\Gamma_I(r+2u), 
\lambda=-2\Gamma_I g,\tilde\theta=-\Re\Theta\sin\phi+\Im\Theta\cos\phi$. 
 Here, $\Re$ and $\Im$ are, respectively, the real and imaginary parts of a 
complex number. This yields $\langle \tilde\theta ({\bf 
x},t)\tilde\theta(0,0)\rangle =2D_\xi\delta({\bf x})\delta(t)$. We 
further equate 
$D_\xi$ with $D$ to recover the noise in Eq.~(\ref{phieq}) above; 
{ advection by a velocity $\bf v$ may be included straightforwardly above in 
Eq.~(\ref{cgle1}), yielding the advective nonlinearity in Eq.~(\ref{phieq}) 
above.} Similar density-dependent CGLE has been derived in different contexts 
recently~\cite{erwin-cgle}. 

\section{Corrections to $\gamma$ for $L \gg L_c$ (without advective nonlinearities)}
As shown in the main text,
\begin{equation}
\gamma_e=\gamma+\overline{\gamma}=\gamma+\frac{ 3\tilde \Omega_3 \langle(\delta 
c)^2\rangle\lambda_2}{D_c}.
\end{equation}

In the regime  $\tau_c/\tau_\times\rightarrow 0$, for $L\gg L_c$,
\begin{equation}
 \langle\delta c^2\rangle \approx \frac{\lambda_2^2 D_1 \tilde\Omega_1^2}{2\pi 
D_c^4 \gamma} \ln 
L.
\end{equation}
Hence,
\begin{equation}
 \overline{\gamma} = 3 \frac{\lambda_2 \tilde\Omega_3}{D_c}\langle (\delta 
c)^2\rangle 
\approx \Gamma \ln L,
\end{equation}
which diverges logarithmically with $L$.

\section{Additional corrections to $\gamma>0$ and $\kappa$ from the advective 
nonlinearities ($\zeta >0$)}

We work in the regime  $\tau_c/\tau_\times\rightarrow 0$. Using 
Eq.~(\ref{genstokes}), we substitute for $v_\alpha$ in Eq.~(\ref{phieq}) 
to obtain a term on the rhs of the latter
\begin{equation}
 \frac{\lambda_1\alpha_0}{\zeta}\left[P_{\alpha\beta}\frac{\partial}{\partial 
x_\gamma} 
\left(\frac{\partial\phi}{\partial x_\gamma}\frac{\partial \phi}{\partial 
x_\beta}\right)\right]\frac{\partial\phi}{\partial x_\alpha}.
\end{equation}

This yields a correction to $\kappa$, given by
\begin{equation}
 \kappa'= \frac{\lambda_1\alpha_0}{\zeta}\int 
\frac{d^2q}{(2\pi)^2}\frac{d\Omega}{2\pi} q^2 \langle |\phi 
({\bf q},\Omega|^2\rangle.
\end{equation}
The correction term is clearly finite for both $L<L_c$ and $L>L_c$. For 
$\lambda_1\alpha_0 >0$, the correction to $\kappa$ (as well as $\kappa_e$) is 
positive and no 
qualitative change is introduced by advection at $O(q^2)$. On the other hand, 
for sufficiently large negative $\lambda_1\alpha_0$, the effective 
$\kappa=\kappa_e+\kappa' < 0$ and the system 
becomes unstable at $O(q^2)$, leading to formation of patterns in the eventual 
steady states.

 We now substitute for $v_\alpha$ in Eq.~(\ref{deneq}) and then 
substitute 
for $\delta c$ in Eq.~(\ref{phieq}). We thus obtain a contribution on the rhs 
of Eq.~(\ref{phieq}) :
\begin{equation}
 \frac{\tilde\Omega_1\lambda_1 \alpha_0\lambda_2}{\zeta D_c^2\nabla^2} 
\left[P_{\alpha\beta} \frac{\partial}{\partial x_\gamma}
\left(\frac{\partial\phi}{\partial x_\gamma}\frac{\partial \phi}{\partial 
x_\beta}\right)\right]\frac{\partial\phi}{\partial x_\alpha}.
\end{equation}

This then yields another correction to $\gamma$, given by

\begin{equation}
 \gamma'= - \frac{\lambda_1\lambda_2\tilde\Omega_1\alpha_0}{\zeta 
D_c^2}\int \frac{d^2 q}{(2\pi)^2}\frac{d\Omega}{2\pi}q^2 \langle |\phi ({\bf 
q},\Omega)|^2\rangle.
\end{equation}
The correction to $\gamma$ is clearly finite for both $L>L_c$ and $L<L_c$. 
Thus, for $\lambda_1\lambda_2\tilde\Omega_1\alpha_0<0$, there are no 
qualitative changes introduced by advection at $O(q^0)$. On the other hand for 
sufficiently large $\lambda_1\lambda_2\tilde\Omega_1\alpha_0>0$, the effective 
active  damping can become negative, yielding instabilities at $O(q^0)$. So far 
we assumed $\gamma>0$ (linear stability). In the linearly unstable case 
($\gamma<0$), it is possible to suppress the linear instability for 
sufficiently large negative $\lambda_1\lambda_2\tilde\Omega_1\alpha_0<0$, such 
that effective active damping can become positive.

%%{\bf TB: can you give the details?}

\end{appendix}

%\end{narrowtext}

%%%%%%%%%%%%%%%%%%%%%%%%%%%%%%%%%%%%%%%%%%%%%%%%%%%%%%%%%%%%%%%%%%%%%%%%%%%%%%%%%%%%

%%%%%%%%%%%%%%%%%%%%%%%%%%%%%%%%%%%%%%%%%%%%%%%%%%%%%%%%%%%%%%%%%%%%%%%%%%%%%%%%%%%%

\begin{thebibliography}{99}
\bibitem{comment-debye} This 
was originally introduced to 
describe 
the depression of the order parameter from its zero temperature maximum due to 
the thermal fluctuations in the 
ordered phases; see, e.g., Ref.~\cite{classicalxy}.
\bibitem{addicomment} We ignore a similar symmetry-allowed $\delta 
c$-dependent feedback term in (\ref{genstokes}) for simplicity.
\bibitem{remark_2} We assume a finite $\gamma$. For small $\tilde\Omega_1$ and 
large 
$\lambda_2$, $L_c\rightarrow\infty$ with a finite $\gamma$.
\bibitem{remark_1} Since we are interested in the long-wavelength limit, 
$O(q^4)$ terms
have been neglected in comparison to $O(q)$ and $(q^2)$.
\end{thebibliography}
\end{document}